\documentstyle[aps,preprint]{revtex}
\tightenlines
\input epsf
\begin{document}
\title{Doping induced metal-insulator transition in two-dimensional 
Hubbard, $t-U$, and extended Hubbard, $t-U-W$, models.}

\author{F.F. Assaad$^{1}$ and  M. Imada$^{2}$ \\
   $^{1}$ Institut f\"ur Theoretische Physik III, \\
   Universit\"at Stuttgart, Pfaffenwaldring 57, D-70550 Stuttgart, Germany. \\
   $^{2}$ Institute for Solid State Physics, University of Tokyo,  \\
    7-22-1 Roppongi,
    Minato-ku, Tokyo 106, Japan.  }

\maketitle

\begin{abstract}
We  show  numerically that the nature of the doping 
induced metal-insulator transition in the two-dimensional Hubbard model 
is radically altered by the inclusion of a term, $W$, which depends
upon a square of a single-particle nearest-neighbor hopping. This result 
is reached by computing the localization length, $\xi_l$,  in the insulating 
state.  At finite values of $W$ we find results consistent with
$\xi_l \sim | \mu - \mu_c|^{ - 1/2} $ where $\mu_c$ is the critical chemical
potential.   In contrast,
$\xi_l \sim | \mu - \mu_c|^{ -1/4}$  for the Hubbard model.
At finite values of $W$, the presented 
numerical results imply  that doping the 
antiferromagnetic Mott insulator leads to a $d_{x^2 - y ^2}$ 
superconductor.  \\
PACS numbers: 71.27.+a, 71.30.+h, 71.10.+x
\end{abstract}

\section{Introduction}
The doping induced metal-insulator transition in the two-dimensional 
Hubbard model is anomalous in the sense that it cannot be understood in terms
of a generic band metal-insulator transition \cite{Imada_95,Imada_rev,Note1}.  
This statement  follows from
numerical work at zero- and finite-temperatures on lattice sizes up 
to $16 \times 16 $.
Zero-temperature quantum Monte-Carlo (QMC) calculations of the
charge susceptibility, $\chi_c  = \partial n  / \partial \mu $,  have shown 
that it behaves as
$ \chi_c \sim | \mu - \mu_c |^{-1/2} $ in the vicinity of the
critical chemical potential $ \mu_c $ at which the transition occurs
\cite{Furukawa_93}.
This result is consistent with $t-J$ model  calculations
\cite{Kohno,Prelovsek}.
In principle, the square root singularity in $\chi_c$ can be
reproduced within a mean-field spin density wave (SDW) approximation 
provided that  perfect nesting is present 
and that  antiferromagnetic order is retained upon shifting the chemical 
potential in the rigid band. When perfect nesting is
absent the SDW approximation yields, as in the case of the generic band
metal insulator transition, a chemical potential independent charge 
susceptibility. This stands in contrast to QMC results, which  for a
{\it small} violation of perfect nesting still show a  square root 
singularity in $\chi_c$ as the critical chemical potential is approached
\cite{Furukawa_93}.
The second piece of information  signaling the anomalous nature of the 
metal insulator transition, is obtained via the numerical calculation
of the localization length, $\xi_l$,  in the insulating state. 
For the two-dimensional Hubbard model one obtains
$\xi_l \sim | \mu - \mu_c | ^{-1/4} $
\cite{Assaad_96}.
In contrast the SDW approximation as well as the generic 
band metal-insulator transition yield a square root singularity  for
$\xi_l$. 
Finally, a QMC calculation of the high frequency Hall coefficient, 
shows no divergence of this quantity in the vicinity of the metal-insulator 
transition \cite{Assaad_95}. 
As opposed to the case of the generic band metal insulator
transition, it's magnitude remains small.

    In a single-particle picture, those numerical observations may be
understood by the occurrence of flat {\it bands} around the $(\pm \pi,0)$,
$ (0, \pm \pi ) $ points in the Brillouin zone. More precisely, 
at the critical 
chemical potential, the exponent obtained numerically from the 
localization length, $\xi_l$,
imposes a $ | \vec{k} |^{4} $ dispersion relation  around  those points.
This flat dispersion relation  required for the understanding of numerical
data, has  important implications. It must originate
from a large wavenumber dependent renormalization of the charge
excitations due to correlation effects since the ordinary van-Hove
singularity does not generate such a pronounced flatness.  
Because of this flatness around the $(\pm \pi,0)$ and 
$(0, \pm \pi)$ points, doped holes predominantly occupy  those
regions.  Thus,  at  low doping the  the regions around the
$(\pm \pi,0)$ and $(0, \pm \pi)$ points in the Brillouin zone govern the low 
energy physics. 
Such strong wavenumber dependent
renormalization may in part be caused by strong scattering of the
quasiparticle between $(\pi,0)$ and $(0,\pi)$ by the antiferromagnetic
fluctuation carrying momentum transfer $(\pi,\pi)$ \cite{Shen97} 
as well as by Umklapp scattering \cite{Note3}.
The flat dispersion thus generated further enhances this scattering
channel in a consistent fashion.
The experimental observation of flat bands, in high-$T_c$ cuprates
is documented in  \cite{Gofron,Dessau,Note2}. 

The occurrence of flat bands suppresses the kinetic energy in the 
metallic state close to the metal-insulator transition. 
This places the metallic 
state as well as the nature of the metal-insulator transition close 
to potential instabilities. In this article, we show that the inclusion
of a term $W$  which depends upon the square of a nearest neighbor hopping
changes radically the  above described properties of the metal-insulator 
transition.  We argue that the low energy physics of the added term, $W$,
is partially contained in the three site terms \cite{Hirsch} obtained in 
a strong coupling expansion (second order perturbation in the hopping) 
of the Hubbard model. 

	Our starting point is the Hubbard model on a square lattice:
\begin{equation}
\label{tU}
      H_{tU} =  -\frac{t}{2} \sum_{\vec{i}} K_{\vec{i}} + 
          U \sum_{\vec{i}}
         (n_{\vec{i},\uparrow}-\frac{1}{2})
         (n_{\vec{i},\downarrow} -\frac{1}{2})
\end{equation}
with the hopping kinetic energy
\begin{equation}
        K_{\vec{i}} = \sum_{\sigma, \vec{\delta}}
   \left(c_{\vec{i},\sigma}^{\dagger} c_{\vec{i} + \vec{\delta},\sigma} +
        c_{\vec{i} + \vec{\delta},\sigma}^{\dagger} c_{\vec{i},\sigma} \right).
\end{equation}
Here,
$c_{\vec{i},\sigma}^{\dagger}$ ($c_{\vec{i},\sigma}$) creates (annihilates) an
electron with {\it z}-component of spin $\sigma$ on site
$\vec{i}$, $n_{\vec{i},\sigma } =  c_{\vec{i},\sigma}^{\dagger}
c_{\vec{i}\sigma}$, and $\vec{\delta} = \pm \vec{a}_x, \pm \vec{a}_y $ where
$\vec{a}_x$, $\vec{a}_y$  are the lattice constants.  
The energy will be measured in units of $t$.  
The interaction we add, is given by:
\begin{equation}
\label{W}
	 H_{W} =  - W \sum_{\vec{i}} K_{\vec{i}}^{2}.
\end{equation}
One motivation of considering this form is that at half-band filling
($\mu  =0$) the Hamiltonian:
\begin{equation}
\label{tUW}
         H_{tUW} =   H_{tU} + H_{W}
\end{equation}
is particle-hole invariant, and the $W-$term has a simple 
Hubbard Stratonovitch transformation.   This has for important consequence 
that the sign problem may be avoided in QMC approaches
at half-filling.
The added $W-$term may have various origins, as pointed out in Ref.
\cite{Assaad_tUW}.
One possible route is to consider the nature of the metallic state near
the Mott transition described above.  The Mott insulating state itself is
characterized by the irrelevance of single-particle processes due to
the opening of a quasiparticle gap. In the insulating state, the 
occurrence of long-range
antiferromagnetic order shows the  relevance of two-particle 
processes  in the particle-hole channel. 
In two-dimensions, the flatness of the charge excitations at small doping
suppress the relevance of single-particle processes and enhances
the importance of two-particle processes.  In contrast to the Mott
insulating state, particle-particle processes now contribute to the
two-particle channel. The form  $H_W$ contains such generic terms of
two-particle processes.

Here, we show the relationship between the $W$-term and the
three site terms obtained in a strong coupling expansion of the  
Hubbard model.  The $W$-term   may be written as 
\begin{equation}
       H_{W} =  H^{(1)}_W + H^{(2)}_W  + H^{(3)}_W + H^{(4)}_W 
\end{equation} 
with 
\begin{equation}
H^{(1)}_W = -8W N
\end{equation} 
where N denotes the number of sites.
\begin{equation}
H^{(2)}_W =  -  W \sum_{\vec{i}, \sigma, \vec{\delta}, \vec{\delta'}}
\left( c_{\vec{i},\sigma}^{\dagger} c_{\vec{i},-\sigma}^{\dagger}
       c_{\vec{i}+ \vec{\delta'},-\sigma}c_{\vec{i}+ \vec{\delta},\sigma}
 + {\rm H.c.} \right)
\end{equation}
\begin{equation}
H^{(3)}_W  = +W \sum_{\vec{i}, \vec{\delta}, \vec{\delta'}, m = -1, 0 , 1 }
\left(  T^{\dagger}_{\vec{i}, \delta', m} T_{\vec{i}, \delta , m} +
        T_{\vec{i}, \delta', m} T^{\dagger}_{\vec{i}, \delta , m} \right)
\end{equation}
and
\begin{equation}
\label{HW4}
H^{(4)}_W = -W \sum_{\vec{i}, \vec{\delta}, \vec{\delta'}}
\left( \Delta^{\dagger}_{\vec{i}, \delta'} \Delta_{\vec{i}, \delta} 
  +  \Delta_{\vec{i}, \delta'} \Delta^{\dagger}_{\vec{i}, \delta} \right).
\end{equation}
Here, $ T^{\dagger}_{\vec{i}, \delta, 1} = c_{\vec{i},\uparrow}^{\dagger}
c_{\vec{i} + \vec{\delta},\uparrow}^{\dagger}$ ,
$ T^{\dagger}_{\vec{i}, \delta,-1} = c_{\vec{i},\downarrow}^{\dagger}
c_{\vec{i} + \vec{\delta},\downarrow}^{\dagger}$ ,
$ T^{\dagger}_{\vec{i}, \delta, 0} = \left(
c_{\vec{i},\uparrow}^{\dagger} c_{\vec{i}+\vec{\delta},\downarrow}^{\dagger} +
c_{\vec{i},\downarrow}^{\dagger} c_{\vec{i}+\vec{\delta},\uparrow}^{\dagger}
\right)/\sqrt{2} $, and
$ \Delta^{\dagger}_{\vec{i}, \delta} = \left(
c_{\vec{i},\uparrow}^{\dagger} c_{\vec{i}+\vec{\delta},\downarrow}^{\dagger} -
c_{\vec{i},\downarrow}^{\dagger} c_{\vec{i}+\vec{\delta},\uparrow}^{\dagger}
\right)/\sqrt{2} $.
In the presence of the Hubbard $U$ with $U>W$,
only $H^{(4)}_W$ should dominate low energy
physics. $H^{(2)}_{W}$ may be neglected since it involves 
processes containing double occupied sites costing an energy $U$. $H_W^{(3)}$
describes triplet pair-hopping. The fact that this term
comes with a positive sign, and that local triplet configurations are not 
favored by the superexchange generated by the Hubbard $U$,  gives us arguments
to conclude that it will not affect the low energy physics. 
Thus, $ H^{(4)}_W$ is the only relevant term. It may be 
rewritten as:
\begin{eqnarray}
H^{(4)}_W & = &  2W \sum_{\vec{i}, \vec{\delta} }
 \left(  \vec{S}_{\vec{i}} \cdot \vec{S}_{\vec{i} + \vec{\delta} } -
\frac{1}{4} \left( n_{\vec{i}} -1 \right) \left( n_{\vec{i} + \vec{\delta} } -1 \right) 
\right)   \nonumber \\
& - & \frac{W}{2} \sum_{\vec{i}, \vec{\delta}  \neq \vec{\delta'}, \sigma }
\left( c_{\vec{i}+\vec{\delta},\sigma}^{\dagger} 
       n_{\vec{i},-\sigma} 
       c_{\vec{i}+\vec{\delta'},\sigma}   
   -   c_{\vec{i}+\vec{\delta},\sigma}^{\dagger} c_{\vec{i},-\sigma}^{\dagger}
       c_{\vec{i},\sigma} 
       c_{\vec{i}+\vec{\delta'},-\sigma}  \right) \nonumber \\
& - & \frac{W}{2} \sum_{\vec{i}, \vec{\delta}  \neq \vec{\delta'}, \sigma }
\left( c_{\vec{i}+\vec{\delta},\sigma}
       \left( 1 - n_{\vec{i},-\sigma} \right)
       c_{\vec{i}+\vec{\delta'},\sigma}^{\dagger}
   -   c_{\vec{i}+\vec{\delta},\sigma} c_{\vec{i},-\sigma}
       c_{\vec{i},\sigma}^{\dagger}
       c_{\vec{i}+\vec{\delta'},-\sigma}^{\dagger}  \right) \nonumber 
\end{eqnarray}
The first term in the above equation is obtained by setting $\delta = \delta'$ 
in  $H_{W}^{(4)}$ and  corresponds to a superexchange. 
Here $\vec{S}_{\vec{i}}$ denotes the spin operator on site $\vec{i}$.  
The second term of the above equation corresponds precisely to the 
three site term \cite{Hirsch}. 
The third term is just the particle hole transform of the three site term. 

The derivation of the relevant terms in $H_W$ via a second
order expansion in $t/U$ is along the same line of idea as the above 
mentioned necessity  to  include two-particle processes.  However,
the strong coupling expansion by itself contains neither the physics of
flat-band formation nor the unusual properties of the
metal-insulator transition.  In that sense, the $W$-term emerges as
relevant processes only after complicated many-body processes of the
flat-band formation and the second order result provides only a starting
point.

In the half-filled case, where  QMC simulations are free of the
sign-problem, the model (\ref{tUW}) has been studied as a function of 
$W/t$ and at fixed value of the Hubbard repulsion, $U/t = 4$ 
\cite{Assaad_tUW}. It has been found that
this model shows a quantum transition at $W_c/t \sim 0.3 $  between an
antiferromagnetic Mott insulator and $d_{x^2 - y^2}$ superconductor. 
Here, we fix  the values of $U$ and $W$ to $U/t =4$ and $W/t < W_c/t$
and study the model as a function of chemical potential. 
The organization of the article is the following. In the next section,
we study the metal-insulator
transition by approaching the  critical point from the insulator side. 
We review the definition of the localization length $\xi_l$ and 
compute it for the parameter set $U/t = 4$ and $W/t = 0.05$ and show that
it diverges as $ | \mu - \mu_c |^{-1/2} $. This is the central result
of the section and stands in contrast to the $W/t = 0$ results:
$\xi_l \sim | \mu - \mu_c|^{-1/4}$. 
We discuss the  relation between the transition
in correlation length exponent between $W/t = 0$ and $W/t > 0.05$  in
terms of the single-particle density of states. 
In Sec.  (\ref{doped}) we consider finite doping. Here, we are confronted to 
a sign problem, which makes it exponentially  hard to reach {\it low} 
temperatures and {\it large} lattice sizes.  We show that at fixed hole 
density and Hubbard repulsion, the $d-$wave pair-field correlations  
increase substantially with  growing values of $W/t$. In contrast no
such increase is observed in the extended $s-$wave channel. 
We equally study the evolution of $n(\mu)$ as  a function of $W/t$. 
The last section is devoted to discussions and conclusions. 

From the technical point of view, our $T=0$ data is obtained with the 
projector QMC (PQMC) algorithm \cite{Koonin,Sandro}.  A numerically 
stable calculation of imaginary time dependent Green functions within this
algorithm is described in Ref. \cite{Assaad_96a}. Finite  temperature  
grand canonical QMC algorithms were equally used \cite{Hirsch85,White}. 
The generalization of those algorithms to efficiently incorporate the
$W-$term is reviewed in \cite{Assaad_tUW}. Dynamical  quantities were 
obtained with the use of the Maximum Entropy method reviewed in 
\cite{Jarrel,Linden}.

\section{Insulator-Metal transition as a function of chemical potential}

We start by studying the  metal-insulator transition by approaching
the critical point from the insulating side. This method has the great
advantage that it allows us to avoid the  sign problem in the QMC method.
This method has already been applied to the case of the single band
two-dimensional
Hubbard model \cite{Assaad_96}. There are implicit assumptions in 
this approach. i)  The
insulating state, is defined by the  vanishing of the Drude weight or 
charge stiffness, $D$. The models we consider are particle-hole symmetric at 
$\mu = 0$ and thus, the critical chemical potential, $\mu_c$,  at which the 
transition occurs is defined by:
\begin{equation}
	D(\mu) = 0 \; \; {\rm for} \; \; |\mu| < \mu_c
\end{equation}
For the range of chemical potentials for which the
Drude weight vanishes, we will assume that the charge susceptibility
equally vanishes:
\begin{equation}
    \chi_c (\mu) \equiv \frac { \partial n (\mu) } { \partial \mu } = 0
\; \; {\rm for} \; \; |\mu| < \mu_c.
\end{equation}
This excludes the possibility of localized states  within the 
charge gap as defined by the vanishing of the Drude weight. 
The occurrence of localized states within the charge gap could 
happen in the presence of disorder. 
ii) Our approach is based on the  knowledge of the single-particle Green
function.  Thus,  we have to assume that the quasiparticle gap, 
$\Delta_{qp}  
= \lim_{N \rightarrow \infty} E_{0}^{N + 1} - E_{0}^{N} $,  satisfies:
\begin{equation}
\Delta_{qp} = \mu_c
\end{equation}
 Here, $N$
denotes the number of sites of the considered square lattice and
$E_0^{N}$ is the ground state energy in the $N$-particle Hilbert space.
In the case 
of an insulator to s-wave superconductor transition,  the  difference 
between $\Delta_{qp}$ and $\mu_c$ is of the  order of the superconducting
gap at the critical chemical potential. In this case, the
insulator-superconductor transition, should
be studied with the two-particle Green function. In the case of a 
transition between an insulator and $d-$wave superconductor  the nodes in the
superconducting gap lead to $\Delta_{qp} = \mu_c$. 
As we will see, in the model of Eq. (\ref{tUW})  the $d-$wave pair-field
correlations dominate over the s-wave pair-field correlations at finite 
doping so that our approach is justified. In the  case of a insulator-metal 
transition where the metallic state is described by a Fermi liquid, the 
relation $\mu_c = \Delta_{qp}$ clearly holds.

With the above assumptions, the localization length we consider is  computed 
as follows. 
The zero-temperature Green function, defined by 
\begin{equation}
\label{Green}
  G(\vec{r}, \omega ) = i\sum_{\sigma} \int {\rm d}t  e^{i \omega t} 
   \langle \Psi_0 |
 T c_{\sigma,\vec{r}}(t)  c_{\sigma,\vec{0}}^\dagger(0) | \Psi_0  \rangle,
  \; \;  {\rm with } \; \;
c_{\sigma,\vec{r}}(t) = e^{i H t} c_{\sigma,\vec{r}} e^{-i H t},
\end{equation}
is real for values of $ |\omega| <  \mu_c $   and satisfies:
\begin{equation}
G(\vec{r},\omega = \mu ) \sim \exp( -|\vec{r}|/\xi_l ) \; \; \; 
     {\rm where}    \; \; \; \xi_l  \sim
| \mu - \mu_c |^{-\nu}  
\end{equation}
Here, $ | \Psi_0  \rangle $ denotes the ground state of the Hamiltonian 
$H \equiv H_{tUW}$ (see Eq. (\ref{tUW})) at half-band filling. 
The correlation length  exponent $\nu $ gives us 
valuable information on the nature of the insulator-metal  transition. 
$\xi_l$ may be interpreted as the localization length of an impurity state
at energy $|\omega| <  \mu_c$.

\begin{figure}
\mbox{}\\[2.0cm]
\epsfbox{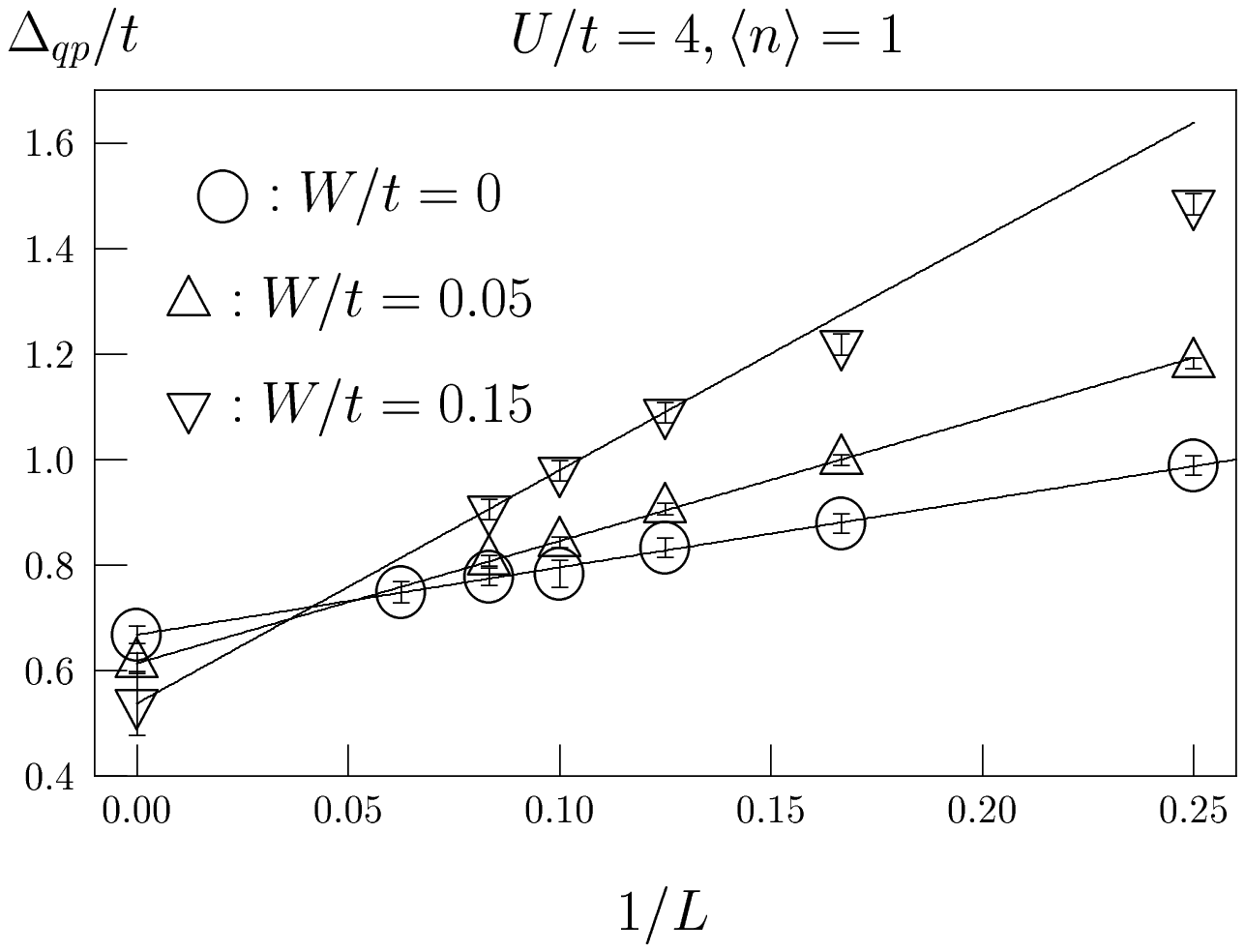}
\mbox{}\\[0.5cm]
\caption[]
{ The quasiparticle gap  $\Delta_{qp} = E_0^{N+1} - E_0^{N} $ as a function
of inverse linear lattice length $L$ at  $U/t=4$ and three values of $W/t$.
Here, $N = L^2$. The solid lines correspond to least square fits to the
form $ a + b/L$. The data points at $1/L = 0$, are the so extrapolated 
values of the quasiparticle gap.  At $W/t = 0.15$  the lattice sizes 
$L = 12, 10$ and $8$ were taken into account for the fit.  
\label{gap.fig} }
\end{figure}

In order to obtain the correlation length exponent, we need an accurate 
evaluation of the critical chemical potential, as well as the  the
single-particle Green function $G(\vec{r},\omega)$ for  $| \omega | < \mu $.
Both those results may be obtained by using the PQMC 
algorithm  described in 
\cite{Assaad_96a} to compute  the imaginary time single-particle 
Green function
\begin{equation}
G(\vec{r}, \tau)  =  \Theta(\tau)
   \frac{ \langle \Psi_0 | \sum_\sigma c_{\vec{r},\sigma}(\tau)
                          c_{\vec{0},\sigma}^{\dagger} |  \Psi_0 \rangle}
        { \langle \Psi_0 |  \Psi_0 \rangle}
         - \Theta(-\tau)
   \frac{ \langle \Psi_0 | \sum_\sigma c_{-\vec{r},\sigma}^{\dagger}(-\tau)
                           c_{ \vec{0},\sigma} |  \Psi_0 \rangle}
        { \langle \Psi_0 |  \Psi_0 \rangle },
\end{equation}
where $c_{\vec{r},\sigma}(\tau) =
e^{ \tau H } c_{\vec{r},\sigma} e^{-\tau H }$.
Here, we use the same notation as in Eq. (\ref{Green}) and 
$\Theta(\tau)$ denotes the Heaviside
function. To obtain an estimate of the quasiparticle gap, $\Delta_{qp}$, we 
fit the  tail of $G(\vec{r}=0, \tau )$ to a single exponential form:
$ \exp ( -\tau \Delta_{qp} )$.  Alternatively, one may compute   $\Delta_{qp}$
by fitting the tail of  $ \sum_{\vec{k}, \epsilon(\vec{k})=0 } G(\vec{k} ,\tau) 
/\sum_{\vec{k}, \epsilon(\vec{k})=0 } $ to the form 
$ \exp ( -\tau \Delta_{qp} )$.  Here, $\epsilon(\vec{k})  = -2 t (
\cos(k_x) +  \cos(k_y) ) $ and $ \epsilon(\vec{k})=0 $ determines the 
non-interacting Fermi  line in half-filled case. 
The so obtained values of are presented in
Fig. \ref{gap.fig} for  $W/t = 0, 0.05 $ and $ 0.15$.  
As argued above, $ \mu_c = \Delta_{qp} $ in the model 
of Eq. (\ref{tUW}).
From Fig. \ref{gap.fig} one notices that size effects become increasingly
important as $W/t$ grows. 

\begin{figure}
\mbox{}\\[2.0cm]
\epsfbox{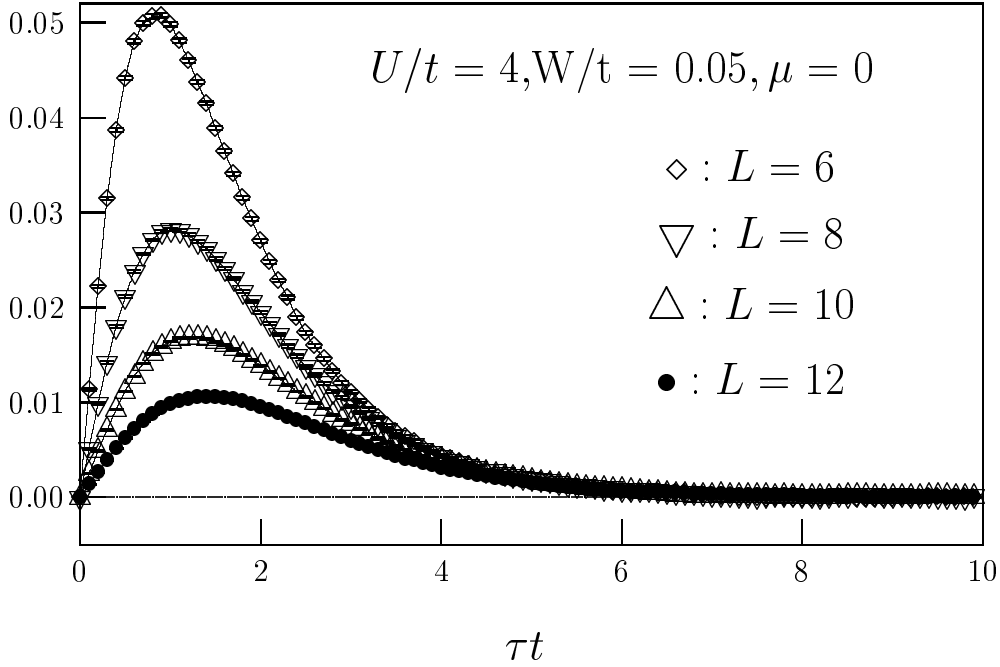}
\mbox{}\\[0.5cm]
\caption[]
{ The single particle Green function at half-filling $\vec{R} = (L/2, L/2) $,
$U/t =4$, $W/t = 0.05$ as a function of imaginary time $\tau$.  
Due to particle-hole symmetry, and the fact that the considered values
of $L$ are even, $ G( \vec{R}, -\tau ) = -G( \vec{R}, \tau ) $.  It is 
interesting to compare this data to the case $W/t = 0$  ( \cite{Assaad_96} ).
At $W/t = 0.05 $, an added electron propagates quicker in imaginary
time than at $W/t = 0 $. 
\label{grL2.fig}}
\end{figure}

For $ | \omega| < \mu_c $, we can efficiently calculate the single-particle 
zero-temperature Green function $G(\vec{r}, \omega)$ through the relation
\begin{equation}
\label{Inter}
     G(\vec{r}, \omega ) = \int_{-\infty}^{\infty} {\rm d}\tau
     G(\vec{r}, \tau) e^{\tau \omega}.
\end{equation}
The Green function $G(\vec{r}, \tau)$
is computed at half-band filling where the sign problem is not present
and the statistical uncertainty does not grow exponentially with lattice
size. However, since we are multiplying the QMC data by the factor
$e^{\tau \omega}$, the statistical uncertainty will grow exponentially with
increasing values of $\tau$ for $\tau \omega > 0$.
For each lattice size $L$,
we have considered the largest distance $\vec{R} = (L/2,L/2)$.
For this distance,
$G(\vec{R}, \tau )$ is plotted in Fig. \ref{grL2.fig}.  Due to
particle-hole symmetry at $\mu =0$, $G(\vec{R}, \tau ) = $
$-G(\vec{R}, -\tau ) $.  

We carry out the imaginary time integration involved in Eq. (\ref{Inter})
in the following way. For  {\it large} positive values of $\tau$, 
$ G(\vec{R}, \tau) \sim \exp(-\tau \Delta_{qp} ) \sum_{\sigma} 
    \langle \Psi_0^N  |  T^{\dagger}_{\vec{R}} c_{\vec{0},\sigma} 
                      T_{\vec{R}}  | \Psi_0^{N+1} \rangle
    \langle \Psi_0^{N+1} |  c_{\vec{0},\sigma}^{\dagger} |  \Psi_0^{N} 
    \rangle $
Here, $ T^{\dagger}_{\vec{R}} c_{\vec{0},\sigma} T_{\vec{R}} = 
c_{\vec{R},\sigma} $. The Hamiltonian (\ref{tUW}) commutes with
$  T_{\vec{R}} $. Since we are working with periodic boundary conditions,
$ T_{\vec{R}}^{2} \equiv 1 $.  Thus, 
$ T_{\vec{R}}  | \Psi_0 >   = \pm  | \Psi_0 >  $,
and  for {\it large} values of $\tau$,  $ G(\vec{R}, \tau)  \sim
\eta G(\vec{0}, \tau) $  with $\eta = \pm 1$.
We can fit the tail of $G(\vec{0}, \tau )$ to the form  
$ \alpha \exp(-\tau \Delta_{qp} ) $ and use this form to estimate the tail of 
$ G(\vec{R}, \tau) $.  The advantage  lies in the fact that it is easier
and more precise to extract the tail from $G(\vec{0}, \tau )$ than from 
$G(\vec{R}, \tau) $.  We now estimate  $G(\vec{R}, \omega ) $ with
\begin{equation}
     G(\vec{R}, \omega ) = \int_{-\tau_1}^{\tau_1} {\rm d}\tau
     G(\vec{R}, \tau) e^{\tau \omega} + 
      \eta \alpha \int_{\tau_1}^{\infty} {\rm d}\tau
         e^{\tau (\omega - \Delta_{qp})} 
   - \eta \alpha \int_{-\infty}^{-\tau_1} {\rm d}\tau  
      e^{\tau ( \omega + \Delta_{qp} ) }.
\end{equation}
Clearly, $\tau_1$ has to be chosen large enough such that the left hand
side of the above equation is $\tau_1$ independent. Numerically, we choose
$\tau_1$ such that for values for $\tau > \tau_1 $,
$ G(\vec{R}, \tau) = \eta G(\vec{0}, \tau) $ within our numerical accuracy. 

\begin{figure}
\epsfbox{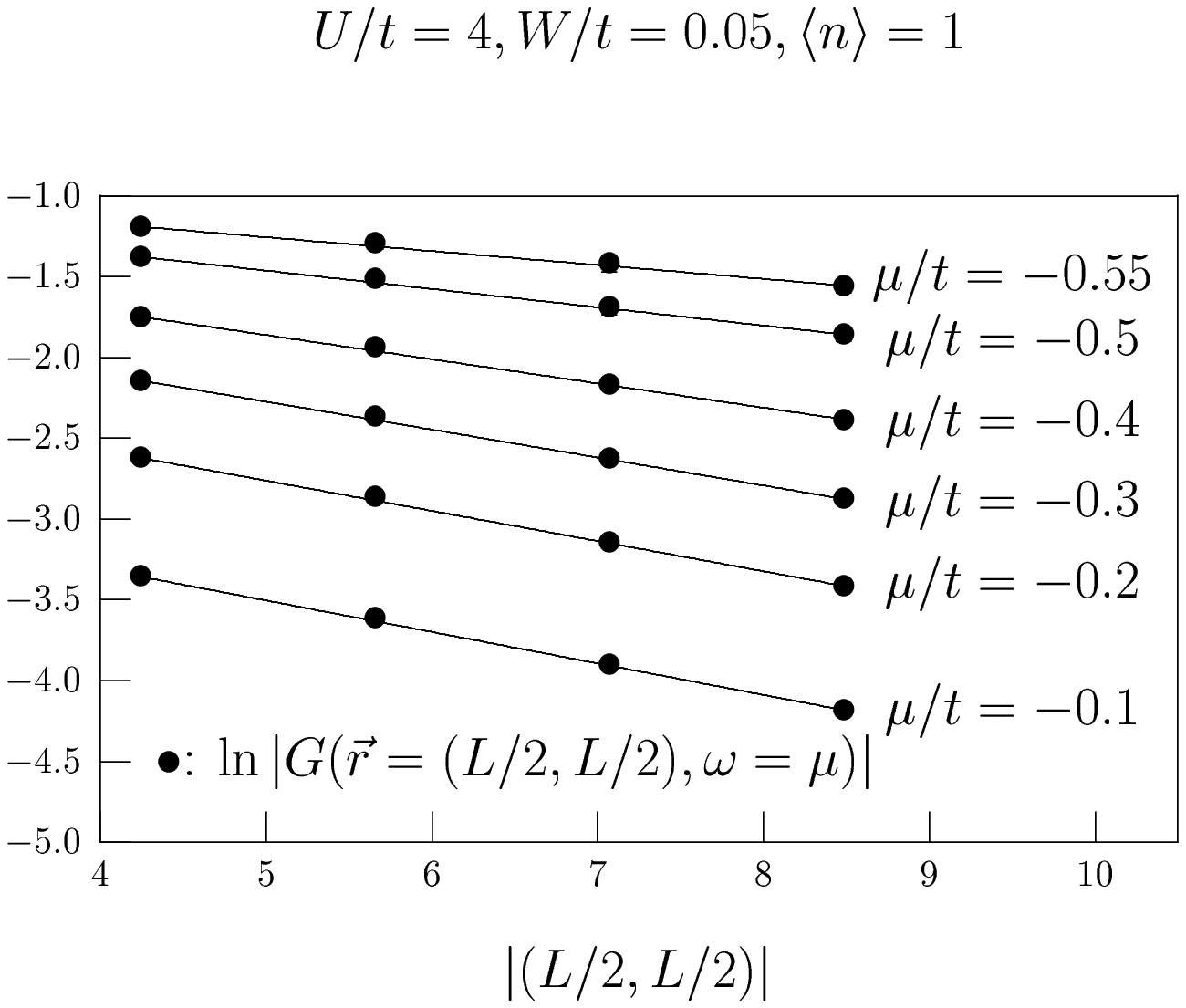}
\mbox{}\\[0.5cm]
\caption[]
{   $ G( \vec{R}= (L/2, L/2), \omega ) $  as obtained from equation
\ref{Inter} as a function of  $L$.  Here we consider the parameter set
$U/t=4$, $W/t = 0.05$  values
of $\omega$ satisfying $\omega < \Delta_{qp} \sim  0.62t$ The
solid lines are least square fits to the form 
$a \exp{ \left(  - L/ \sqrt{2} \xi_l \right) } $.
\label{loc.fig} } 
\end{figure}

\begin{figure}
\epsfbox{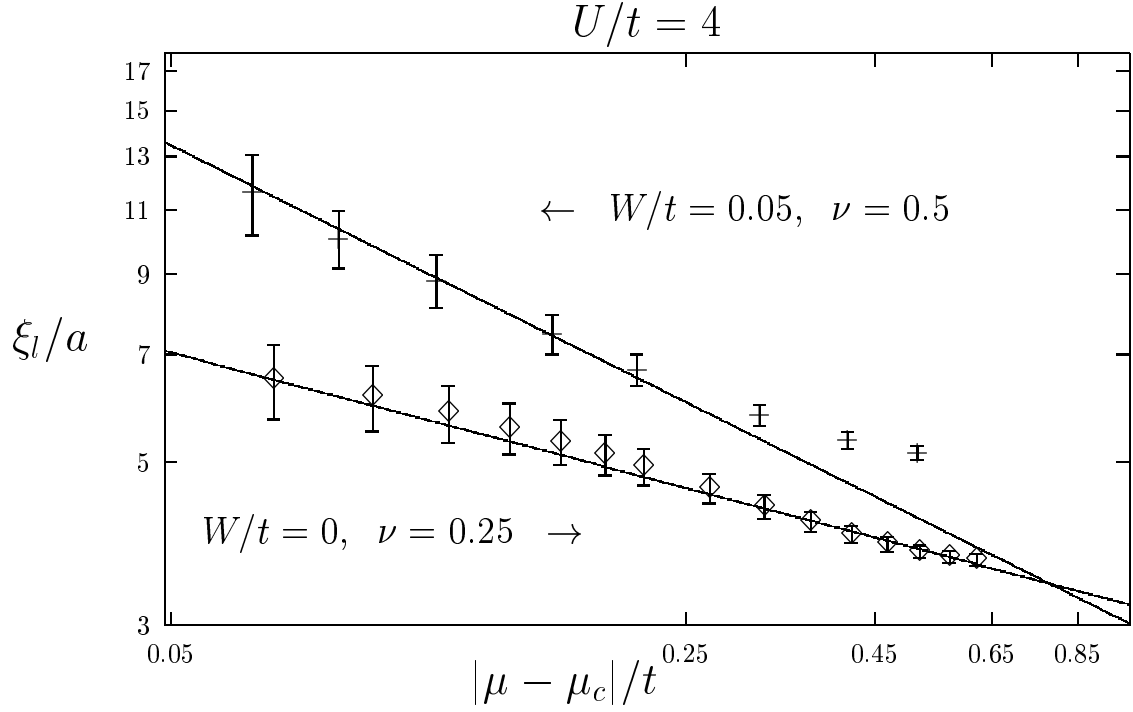}
\mbox{}\\[0.5cm]
\caption[]
{The localization length as a function of $ |\mu - \mu_c| $. The data at
$W/t = 0$, $U/t = 4$ is reproduced from Ref. \cite{Assaad_96}.  The
data at $W/t = 0.05$, $U/t=4$ is obtained from Fig. (\ref{loc.fig}).  The
solid lines correspond to the form 
$ | \mu - \mu_c | ^{-\nu} $ for  $ \nu = 0.5 $ and $ \nu = 0.25 $.  The
data is consistent with 
$ | \mu - \mu_c | ^{-1/2} $  at $W/t = 0.05$ and
$ | \mu - \mu_c | ^{-1/4} $  at $W/t = 0$.
\label{xi_tot.fig}}
\end{figure}

The so obtained values of  $G(\vec{R}, \omega = \mu )$ are plotted in 
Fig. \ref{loc.fig} at $U/t = 4$ and $W/t = 0.05$.
for $L = 6$ to $L = 12$  lattices.  One may fit the data to the form
$\exp (-|\vec{R}|/\xi_l)$ to obtain the localization length $\xi_l$.
The so obtained localization length is plotted versus $ | \mu - \mu_c | $ 
on a log-log scale in Fig. \ref{xi_tot.fig}. For comparison, we have
included our estimate of
the localization length at $U/t = 4$ and $W/t = 0$. The central result
of this section is the fact that the numerical data are consistent  with
\begin{eqnarray}
    \xi_l \sim | \mu - \mu_c |^{-1/4} \; \; {\rm at} \; \; 
U/t = 4, W/t & = & 0 \nonumber \\
    \xi_l \sim | \mu - \mu_c |^{-1/2} \; \; {\rm at} \; \; 
U/t = 4, W/t & = & 0.05
\end{eqnarray} 
We also note that the localization length grows at fixed $\mu$ with 
increasing $W/t$. At $W/t = 0.15$ growing size effects rendered the 
determination of the localization length difficult for the
studied system sizes. 

\begin{figure}
\mbox{}\\[-1.0cm]
\epsfbox{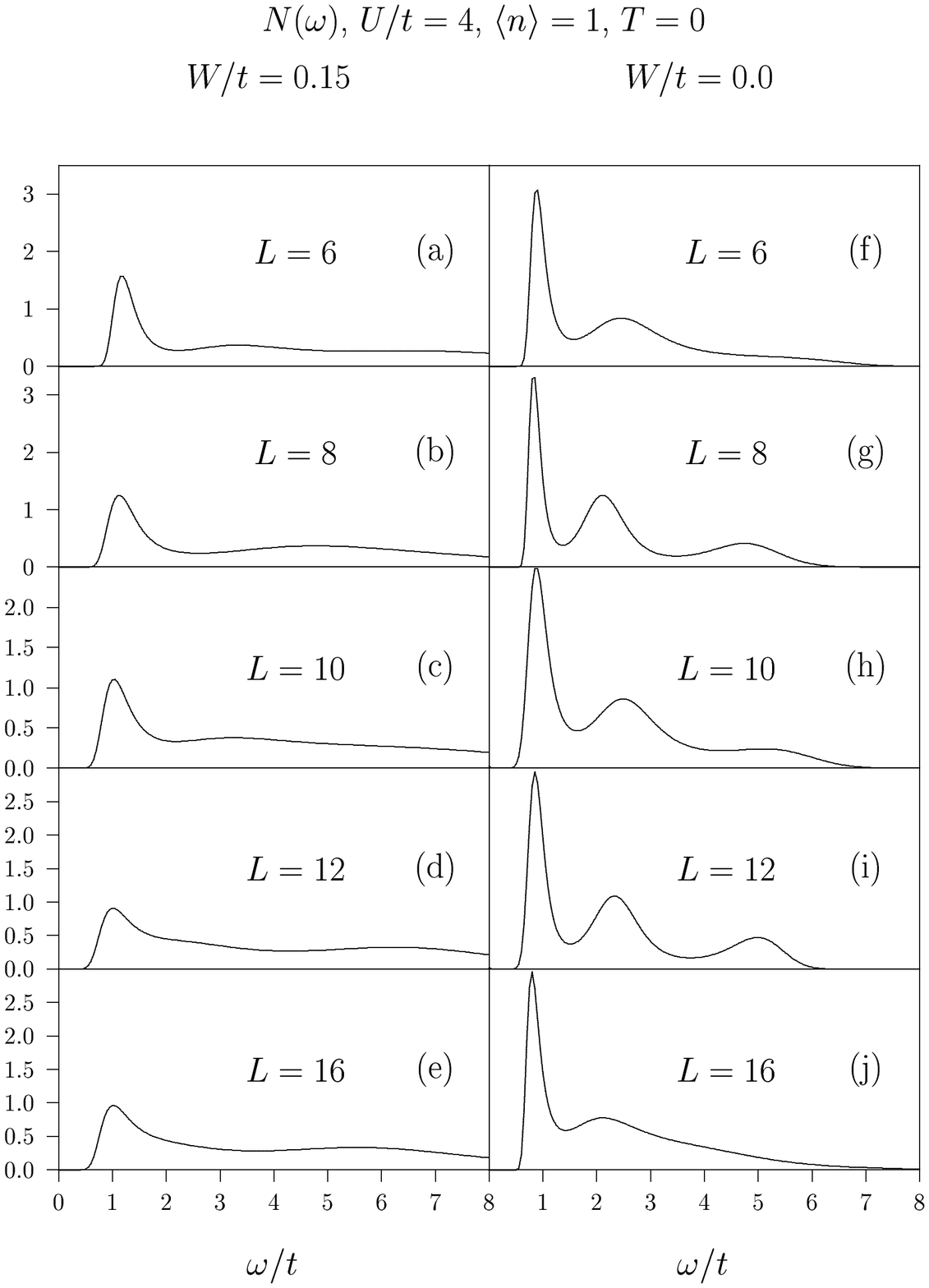}
\newpage
\epsfbox{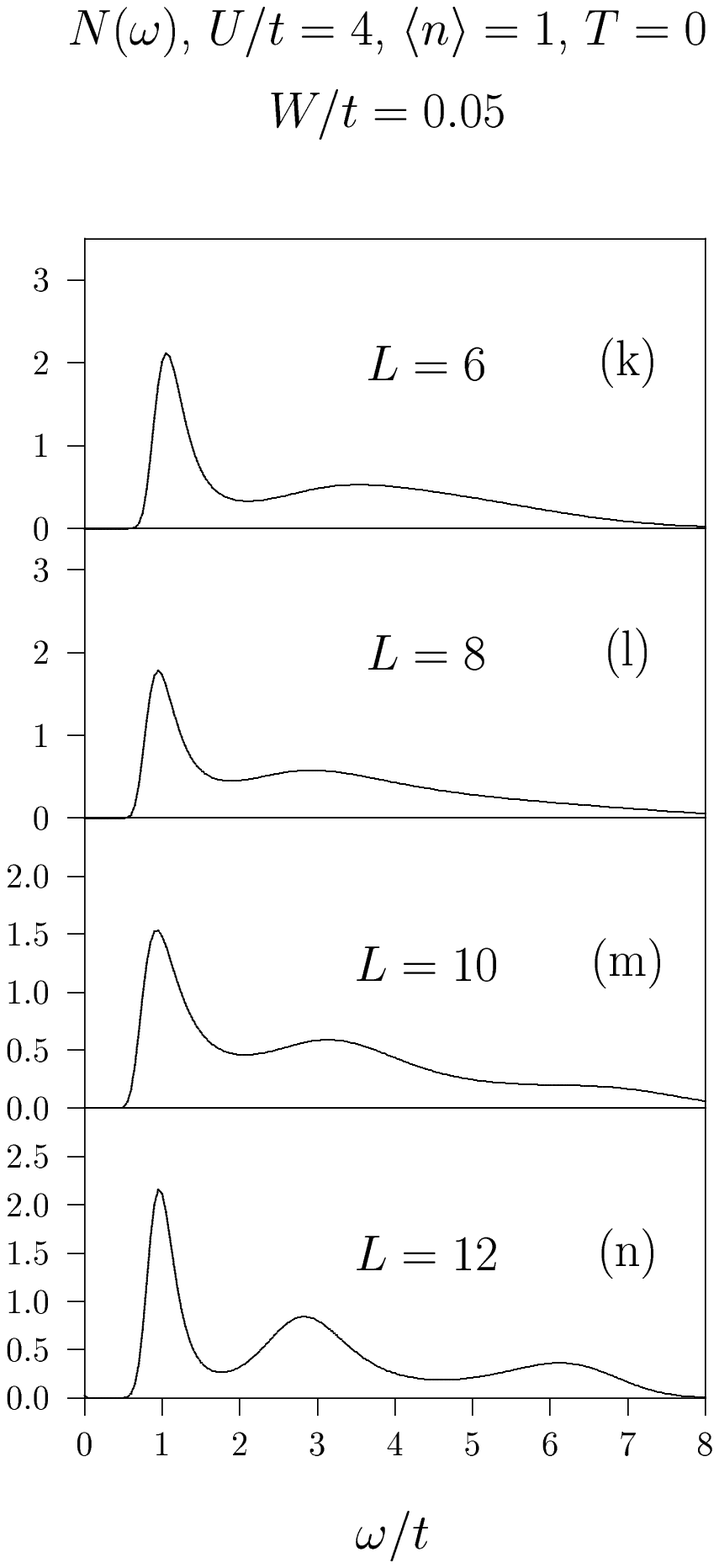}
\caption[]
{$T=0$ density of states, $N(\omega)$ at half-band filling 
(i.e. $\mu = 0$ ) for $U/t = 4$  and (a) to (e) $W/t = 0.15$, 
(f) to (j)  $ W/t = 0$ and (k) to (n) $ W/t = 0.05$.
\label{nom.fig}}
\end{figure}

As mentioned in the introduction,  the  correlation length exponent 
$\nu = 1/4$ at $W/t = 0$ may be understood in a single-particle picture
by the existence of a 
$|\vec{k}|^{4}$ dispersion relation around the $(\pm \pi , 0)$ 
$(0,\pm \pi)$  points in the  Brillouin zone. This 
very flat dispersion relation should lead to a high density of
states, 
$N(\omega) \equiv \frac{1}{N}\sum_{ \vec{k} } {\rm Im} G(\vec{k},\omega)$, 
for frequencies close to the quasiparticle gap. In comparison, 
at finite values of $W/t$ where we obtain the correlation length 
exponent $\nu = 1/2$, we expect no anomalous enhancement of 
$N(\omega)$ at frequencies equal to the quasiparticle gap. In fact, for the 
generic metal-insulator transition for two-dimensional systems \cite{Note1},  
for which one obtains the exponent $\nu = 1/2$, $ N(\omega) $ 
remains a finite constant  until it jumps to zero 
at  the band-edge. 
In order to check this idea, we have computed the one-electron density 
of states:
$ N (\omega) $ at half-band filling. At zero-temperature and $\tau t > 0$, 
$N(\omega)$ is related to the imaginary time QMC data via
\begin{equation}
      G(\vec{r}=\vec{0},\tau) = \frac{1}{\pi} \int_{0}^{\infty} 
  {\rm d} \omega e^{-\tau \omega} N(\omega ).
\end{equation}
$N(\omega)$ is extracted from the above equation with the use of the 
Maximum-Entropy method. 
Fig. \ref{nom.fig} plots $N(\omega)$ 
for lattice sizes ranging from $L=6$ to $L=16$ at $U/t = 4$ and 
for both $W/t =0$ ( Figs. \ref{nom.fig}f to \ref{nom.fig}j )
and $W/t = 0.15$ ( Figs. \ref{nom.fig}a to \ref{nom.fig}e ).
Figs.  \ref{nom.fig}k to \ref{nom.fig}n plots the same data but
for  $W/t = 0.05$.
It is clear that as $W/t$ is enhanced,
the large density of states present at 
$\omega \sim \Delta_{qp}$ vanishes.

\section{Finite Doping}
\label{doped}

We now consider finite doping. Here, we are confronted
with a sign problem  which leads, for a given accuracy, to an exponential 
increase of CPU time as the lattice size and inverse temperature grows. 

\begin{figure}
\mbox{}\\[-2.0cm]
\epsfbox{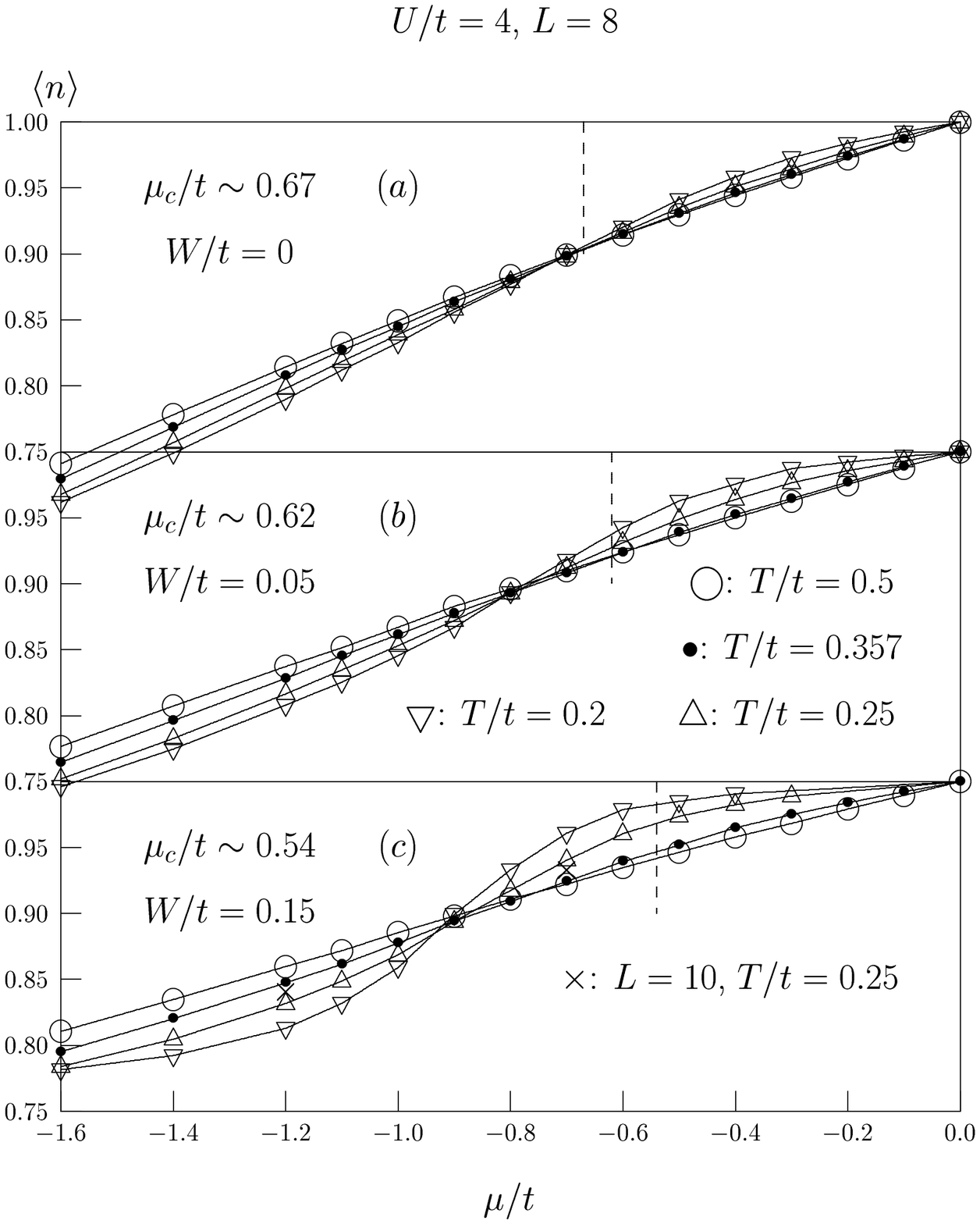}
\caption[]
{Number of particles per site, n, as a function of chemical potential
for  temperatures ranging from $T= 0.2 t$ to $T  = 0.5 t$.  The dashed 
vertical line corresponds to the  quasiparticle gap  at $T=0$ and in the 
thermodynamic limit. This data is obtained from Fig. \ref{gap.fig}.
We consider an $8 \times 8 $ lattice, $U/t = 4$ and 
(a)  $W/t = 0$, (b)  $W/t = 0.05$  and (c)  $W/t = 0.15$.  In Fig. (c)
we have included some data points for a $10 \times 10$ lattice. 
\label{chem.fig}}
\end{figure}

We first consider the particle number as a function of chemical potential
at $U/t = 4$ for $W/t = 0, W/t =  0.05$ and $W/t = 0.15$ on an 
$ 8 \times 8 $ lattice in a temperature window $T = 0.5 t $ to $T = 0.2 t$. 
The data is plotted in Fig. (\ref{chem.fig}).
Finite size effects grow as a function of $W/t$. This may be seen for example
in Fig. \ref{gap.fig} where it becomes increasingly hard to extrapolate to
the thermodynamic limit for {\it large } values of $W/t$. 
To give the reader a feeling of the size effects involved, we have 
plotted at $W/t = 0.15$ and $T = 0.25t$, some data for an $L=10$ lattice
(see Fig. \ref{chem.fig})
For the discussion of the data, we will omit the size effects in the 
considered temperature range. 
The dashed lines correspond to the value of the quasiparticle
gap obtained in the zero-temperature and thermodynamic limits.  
(see Fig.\ref{gap.fig}).  As discussed above, we expect for our model
$\Delta_{qp} = \mu_c$. Hence, at $T=0$, $ n(\mu) = 1 $ for $ \mu > \mu_c$
and $ n(\mu) < 1  $ for $ \mu < \mu_c$.  From this construction,  one
may see that temperature effects are much more pronounced at $W/t =0$ than
at $W/t = 0.15$.  
This result is easily understood within a  single-particle picture 
mentioned in the introduction. The flatness of the dispersion relation
signaled by the correlation length exponent $\nu = 1/4$ at $W/t = 0$ leads 
to large temperature effects in $n(\mu)$. The transition to the description 
of the insulator-metal transition with exponent $\nu = 1/2$, leads to
the suppression  of large temperature effects in $n(\mu)$. At
$W/t = 0$ it was possible to pin down the exponent of the charge susceptibility,
$\chi_c \sim | \mu - \mu_c | ^{-1/2} $ 
only with the use of zero-temperature algorithms such as the PQMC 
\cite{Furukawa_93,Kohno} . 
As argued above,  at finite values of $W$, temperature effects are reduced. 
Nevertheless,  it is still  difficult to pin down the 
charge susceptibility exponent from finite temperature  data.

\begin{figure}
\mbox{}\\[-2.0cm]
\epsfbox{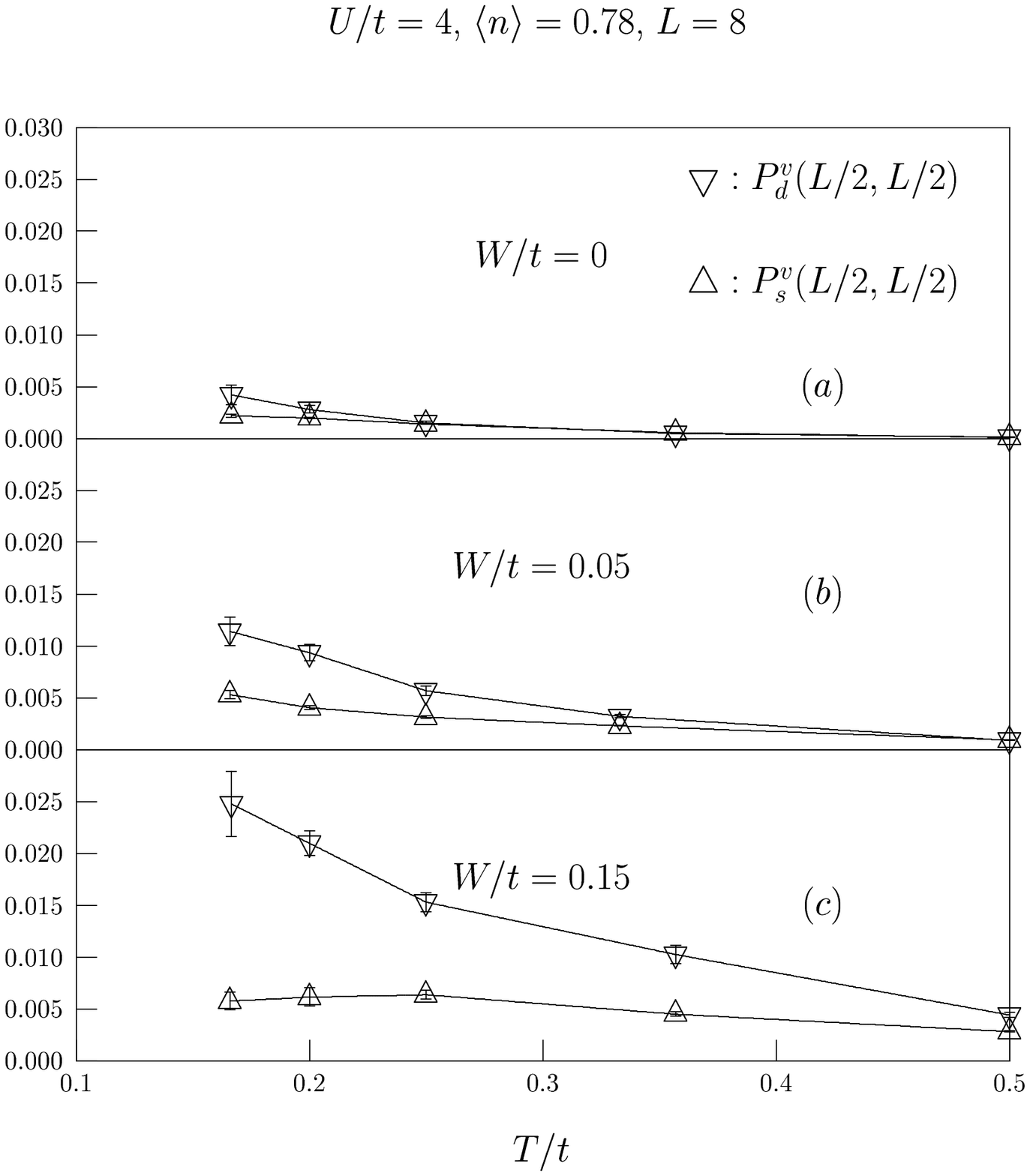}
\mbox{}\\[0.5cm]
\caption[]
{Vertex contribution to the pair-field correlations at the largest distance
on an $L=8$ lattice in both the $d-$wave and extended $s-$wave channel. The
data is plotted versus temperature. 
Here we fix the  the particle number to $\langle n \rangle  =0.78 $, the 
Hubbard repulsion to $U/t = 4$  and consider  (a) $W/t = 0$, (b) $W/t = 0.05$
and (c) $W/t = 0.15$ \label{pair.fig} } 
\end{figure}

We conclude this section by giving some numerical evidence, that the
cause of the transition in the nature of the insulator-metal transition is
related to the enhancement of $d_{x^2 - y^2}$ pairing correlations  in the
metallic state. The vertex contribution to the  equal
time pair-field correlations is given by:
\begin{equation}
 P_{d,s}^{v} (\vec{r}) = P_{d,s} (\vec{r})  -
   \sum_{\sigma,\vec{\delta}, \vec{\delta}' }  f_{d,s}(\vec{\delta})
        f_{d,s}(\vec{\delta}')
\left(
 \langle c^{\dagger}_{\vec{r},\sigma} c_{\vec{\delta}',\sigma}   \rangle
 \langle c^{\dagger}_{\vec{r}+\vec{\delta},-\sigma} c_{\vec{0},-\sigma} \rangle
+ \langle c^{\dagger}_{\vec{r},\sigma} c_{\vec{0},\sigma} \rangle
 \langle c^{\dagger}_{\vec{r}+\vec{\delta},-\sigma}
         c_{\vec{\delta}',-\sigma} \rangle \right)
\end{equation}
where
\begin{equation}
P_{d,s} (\vec{r}) = \langle \Delta_{d,s}^{\dagger}(\vec{r})
\Delta_{d,s}(\vec{0}) \rangle
\end{equation}
with
\begin{equation}
\Delta_{d,s}^{\dagger}(\vec{r})  =
\sum_{\sigma,\vec{\delta}}  f_{d,s}(\vec{\delta})
\sigma c^{\dagger}_{\vec{r},\sigma}
c^{\dagger}_{\vec{r} + \vec{\delta},-\sigma}.
\end{equation}
Here, $f_{s}(\vec{\delta}) = 1$ and $f_{d}(\vec{\delta}) = 1 (-1)$
for $\vec{\delta} = \pm \vec{a_x}$ ($\pm \vec{a_y})$.
Per definition, $ P_{d,s}^{v}  (\vec{r}) \equiv 0 $ in the absence of
interactions.
Fig. \ref{pair.fig}  plots 
$ P_{d,s}^{v}  (L/2,L/2) $ at $ \langle n \rangle = 0.78 $,
and $U/t = 4$ as a function of temperature.  We consider  an $ 8 \times 8$
lattice with  $W/t = 0$, $W/t = 0.05$ and $W/t = 0.15$.  As $W/t $ 
grows, there is a substantial increase in $P_{d}^{v} (L/2,L/2)$.  This stands
in contrast to the signal obtained in the extended $s-$wave channel where
virtually no enhancement as a function of $W/t$ is seen. 

\section{Discussion and Conclusions}
We have given numerical evidence showing that the nature of the metal insulator
transition in the two-dimensional Hubbard model is radically altered by
the inclusion of the $W-$term. The low energy physics contained in this term
was argued to be described by the three site terms obtained in a
strong coupling expansion of the Hubbard model. The presented data support
the interpretation that the  change in the nature of the doping induced 
metal insulator transition as a function of $W/t$ is related to the onset of
$d-$wave superconductivity in the metallic state at finite $W/t$. 
The critical value of $W$, $W^{\mu}_c/t $, at
which the nature of the metal-insulator transition changes is uncertain. 
We can  at present only give an upper-bound: 
$W^{\mu}_c/t < 0.05 $ for $U/t = 4$. 

Our result are best described under the assumption of hyperscaling  
\cite{Imada_95,Imada_rev}
where it is postulated that the singular part of the free energy 
scales as  $f_{s} ( \Delta ) = \Delta^{\nu \left( d + z  \right) }$  and 
that there is a single relevant length scale $\xi$.
Here, $\Delta \equiv  | \mu - \mu_c |  $ is related to the length
$\xi \sim \Delta^{-\nu}$,  $d$ is the dimensionality and $z$ the
dynamical exponent. 
Under the above assumptions, the charge susceptibility $\chi_c$, charge
stiffness $D$, and localization length $\xi_l$, satisfy: 
\begin{equation}
\label{Scale}
     \xi_l \sim \Delta^{-\nu}, \; \;
     \chi_c  \sim \Delta^{\nu (d-z)}, \; \;
     D \sim \Delta^{\nu (d + z - 2)}, \; \;
\end{equation}
Since the control parameter, $\Delta$, 
corresponds to the chemical potential, one obtains the 
additional constraint $\nu z = 1$ as well as 
$\delta \sim \Delta^{\nu (d + z) -1} $, $\delta$ being the 
doping concentration. Thus, in the case of a doping induced  metal-insulator
transition,  there is only one free parameter. The generic band metal-insulator
transition does satisfy the above assumption and  belongs to the $z =2$,
$ \nu = 1/2 $ universality class. 
We also note that the doping induced metal-superfluid transition for
two-dimensional bosons, is described by hyperscaling with exponents:
$\nu  = 1/2$ and $z = 2$ \cite{Fisher89}.
For the two-dimensional Hubbard
model at $U/t = 4$, we have two numerical results, 
the localization length as well
as the charge susceptibility, which are consistent with the above
scaling  relations provided that $\nu = 1/4 $ and $z = 4$.
At $W/t =  0.05$ and $U/t =4$,  our results for the localization length,
$ \xi_l \sim | \mu - \mu_c |^{-1/2}$ imply that $ \nu = 1/2 $.   
Furthermore, from our finite temperature results for $n(\mu)$, it
is plausible to assume that  $\chi_c \sim |\mu - \mu_c |^{0}$ in the 
zero-temperature 
and thermodynamic limits as would be expected if the hyperscaling
assumption is satisfied. 
For the confirmation the hyperscaling  assumption, 
the calculation of the charge stiffness at
$T=0$ as a function of doping and $W/t$ is crucial. 
The prediction that $D \sim \delta^2 $ at $W = 0$ and that 
$D \sim \delta $ at $W > W_c^{\mu}$  remains for further studies.

The above analysis suggests that the metal-insulator transition
characterized by the dynamical exponent $z=4$ is destabilized by a finite
value of $W$ to the superconductor-insulator transition with $z=2$.  When
the dynamical exponent $z$ changes from $z=4$ to $z = 2$ by inclusion 
of a finite
amplitude of the $W$-term, the thus obtained wider dispersion relation
causes a strong kinetic energy gain. 
This may be the origin of the instability of the
incoherent metals with $z=4$ to the superconducting state with $z=2$.  A
flat band structure itself could be mimicked by a simple modification of
the original band structure (e.g the inclusion of next-nearest neighbor
transfer). 
However, to understand the instability due to the change in
the universality class and the origin of the release of the suppressed
coherence (i.e. $D \sim \delta^2$), it is crucial to notice 
that the $z=4$ metals are a consequence
of strong correlation effects and have to be constructed not from simple
band electrons but from strongly renormalized charge excitations around the
$(\pm\pi, 0)$ and $(0,\pm\pi)$ points in the Brillouin zone
\cite{Imada_rev,Imada97}. 

	 We have previously shown that at $T=0$, 
$\langle n \rangle=1$, and $U/t = 4$,  $W$ drives the system from an 
antiferromagnetic Mott insulator to a $d_{x^2 - y^2}$ superconductor. This
quantum transition occurs at
$W_c/t \sim  0.3$.   In the superconducting state at half-band filling,
charge and spin degrees of freedom have been studied and show rich 
physics \cite{Assaad_tUW}.  The data shows the extreme compatibility 
between antiferromagnetic fluctuations and $d$-wave superconductivity. 
The relation between this state and the state obtained by doping the  
antiferromagnetic Mott insulator  at $U/t =4$, $ W^{\mu}_c < W < W_c $ 
is intriguing.  High-$T_c$ cuprates are obtained by doping the Mott
insulator. Some aspects of the experimental data  
seem to show the remnant of extreme compatibility between magnetism
and $d-$wave superconductivity. 
In nuclear magnetic resonance studies of  $Y Ba_2 Cu_4 O_8$ and
$Hg Ba_2 Ca_2 Cu_3 O_{8 + \delta}$, 
$ \; T_{2G}$ of $Cu$-sites, which measures predominantly
${\rm Re} \chi ( \vec{q} = (\pi,\pi), \omega = 0 ) $ continues to grow 
monotonically with decreasing temperature  
even below the temperature scale at which 
$1/T_1 T$  starts decreasing \cite{Itoh}. Although this feature is not
clear cut in the single-layer-type mercury compounds
\cite {Itoh97} the tendency of stronger
growth of $1/T_{2G}$ than $1/T_1 T$ above $T_c$ seems to be universal. 
An  extreme case of this feature is seen in our numerical calculations 
in the superconducting state at $ \langle n \rangle = 1 $ where  in the 
low temperature limit, $1/T_1 T $ scales  to zero while 
${\rm Re} \chi ( \vec{q} = (\pi,\pi), \omega = 0 ) $ diverges. 
The most remarkable feature is that the
dynamical spin
structure factor, $S(\vec{q},\omega)$,  at  $\vec{q} = (\pi,\pi)$ has
has a peak structure with diverging weight at finite frequencies
\cite{Assaad_tUW}. 
In $La_{2-x} Sr_x Cu O_4$, 
$S(\vec{q},\omega)$,  below $T_c$ shows a sharp
peak in momentum  space centered at the incommensurate wave-vector,
$\vec{Q}_\delta = (\pi,\pi) + \delta (\pi,0) $ 
and at relatively high energies: $\omega > 7 meV $ \cite{Mason}.
This indicates that, the magnetic length scale at relatively high
energies becomes
large ( $ > 50 {\AA} $ ) as the temperature is lowered below $T_c$.
In $Y Ba_2 Cu_3 O_{7-x} $, a  sharp peak in frequency at $30-40$ $meV$ 
is observed  in the superconducting phase \cite{Rossat-Mignod,Fong}.
These  features again share some similarity  with 
the $t-U-W$ model at $\langle n \rangle = 1$ in the superconducting state. 
At temperatures below the Kosterlitz-Thouless transition,
spectral weight  in $S( \vec{q}, \omega) $ is transfered from
low frequencies to higher frequencies.  In the low temperature limit,
$S(\vec{q} = (\pi,\pi), \omega )$ shows a sharp peak in momentum space 
at finite frequencies. This peak is associated with the coherent part of 
the spin response since it is related to the long distance power-law 
decay of the equal time spin-spin correlations \cite{Assaad_tUW}. 

In conclusion, the $t-U-W$ model at finite values of $W/t$ appears 
to be a promising  minimal model for high-$T_c$ superconductors.
Clearly, further  studies for the complete understanding of physical 
properties of the model 
as function of $W$, doping and temperature are required. 

We thank D.J. Scalapino for many stimulating discussions as well  as
H. Tsunetsugu and A. Muramatsu. The  numerical calculations were carried 
out on the FACOM VPP500 at the Supercomputer Center of the Institute for
Solid State Physics, Univ. of Tokyo.  This work was financially supported by 
a Grant-in-Aid for `Research for the Future' program under the project 
number
JSPS-RFTF97P01103,  by a Japan-Germany joint research project 
from the Japan Society for  the Promotion of Science  as well as by 
a Grant-in-Aid for Scientific Research on the Priority  Area `Anomalous
Metallic State near the Mott Transition' from the Ministry  of Education
Science and Culture Japan. 
F.F.A.  thanks the Swiss National Science  foundation for partial 
financial support under the grant number 8220-042824.

\newpage

\newpage

\newpage

\newpage

\newpage

\newpage


\begin{thebibliography}{99}
\bibitem{Imada_95} M. Imada,  J. Phys. Soc. of Jpn. {\bf 64}, 2954 (1995).

\bibitem{Imada_rev} M. Imada, A. Fujimori and Y. Tokura.
Metal-Insulator Transitions, to be published in  Rev. Mod. Phys.

\bibitem{Note1} Our working definition
of the generic band metal-insulator transition is the one realized for
free fermions, $ H_0 = \sum_{\vec{k},\sigma }  \frac{\vec{k}^2}{2m}
c^{\dagger}_{\vec{k},\sigma} c_{\vec{k},\sigma}  $
at zero chemical potential $\mu = 0$.

\bibitem{Furukawa_93}  N. Furukawa and M. Imada, J. Phys. Soc. Jpn. {\bf 62},
2557, (1993).

\bibitem{Kohno} M. Kohno,
Phys. Rev. B  {\bf 55}, 1435, (1997).

\bibitem{Prelovsek} J. Jaklic and P.  Prelovsek,
Phys.  Rev.  Lett.  {\bf 77}, 892,  (1996).


\bibitem{Assaad_96} 
F.F. Assaad and M. Imada, Phys. Rev. Lett. {\bf 76}, 3176, (1996).

\bibitem{Assaad_95} 
F.F. Assaad and M. Imada, Phys. Rev. Lett. {\bf 74}, 3872, (1995). 


\bibitem{Shen97} 
Z.-X. Shen and J. R. Schrieffer, Phys. Rev. Lett. {\bf 78}, 1771, (1997).

\bibitem{Note3}
In the presence of long-range antiferromagnetic order, the vector 
$ (\pi, \pi) $  belongs to the reciprocal lattice since the Brillouin
has been folded. Growing  antiferromagnetic 
correlations will give some precursor of this Umklapp scattering.

\bibitem{Gofron}
K. Gofron, el al. Phys. Rev. Lett. {\bf 73}, 3320, (1994).

\bibitem{Dessau} 
D.S. Dessau et al. Phys. Rev. Lett.  {\bf 71}, 2781, (1993).
Z.-X. Shen and D.S.Dessau,Phys. Rep. {\bf 253}, 1, (1995). 

\bibitem{Note2}
The flat bands observed experimentally, are much flatter than the usual
Van-Hove singularity.  In Ref. \cite{Gofron} it has been argued
that they yield a powerlaw singularity in the density of states. 

\bibitem{Assaad_tUW} F.F. Assaad, M. Imada and D.J. Scalapino, Phys. Rev. Lett.
{\bf 77}, 4592,  (1996). Phys. Rev. B, in press. 

\bibitem{Hirsch} J.E. Hirsch, Phys. Rev. Lett {\bf 54}, 1317, (1985).


\bibitem{Koonin} G. Sugiyama and S.E. Koonin, Annals of Phys.{\bf 168}
(1986) 1.

\bibitem{Sandro} S. Sorella, S. Baroni, R. Car, And M. Parrinello,
Europhys. Lett. {\bf 8} (1989) 663.
S. Sorella, E. Tosatti, S. Baroni, R.
Car, and M. Parinello, Int. J. Mod. Phys. B{\bf 1} (1989) 993.

\bibitem{Assaad_96a} 
F.F. Assaad and M. Imada, J. Phys. Soc.  Jpn. {\bf 65},189, (1996). 

\bibitem{Hirsch85} J.E.Hirsch, Phys. Rev. B {\bf 31}, 4403, (1985).

\bibitem{White} S.R. White et al.  Phys. Rev. B{\bf 40}, 506, (1989).

\bibitem{Jarrel} M. Jarrel and J.E. Gubernatis, 
Physics Reports, {\bf 269}, (1996) 133

\bibitem{Linden} W. von der Linden, 
Applied Physics A{\bf60},   (1995), 155.


\bibitem{Fisher89}
M.P.A. Fisher, P.B. Weichman, G. Grinstein and D.S. Fisher: Phys. Rev. B
{\bf 40} (1989) 546.

\bibitem{Imada97} M. Imada,
In Computer Simulation Studies in Condensed matter Physics X, ed.
by D.P.Landau et al. (Springer Verlag, Heidelberg, Berlin, 1997).

\bibitem{Itoh}  Y. Itoh et al. 
Y. Itoh, H. Yasuoka, Y. Fujiwara, Y. Ueda, T. Machi, I.
Tomeno, K. Tai, N. Koshizuka and S. Tanaka, J. Phys. Soc. Jpn., {\bf 61}, 
1287, (1992).
M.H. Julien et al. Phys. Rev. Lett {\bf 76}, 4238, (1996).

\bibitem{Itoh97} Y. Itoh et al. preprint. 

\bibitem{Mason} T.E. Mason, A. Schr\"oder, G. Aeppli, H.A. Mook, and S.M.
Hayden, Phys. Rev. Lett. {\bf 77}, 1604, (1996).

\bibitem{Rossat-Mignod} J. Rossat-Mignod et al. Physica  (Amsterdam) 
{\bf 185C - 189C}, 86, (1991).

\bibitem{Fong} H.F. Fong et al.  Phys. Rev. Lett. {\bf 75}, 316, (1995).
H.F. Fong et al. Phys. Rev. Lett. {\bf 78}, 713, (1997). 

\end{thebibliography}
\end{document}